\documentclass[preprint,showpacs,preprintnumbers,superscriptaddress,amsmath,amssymb, prb]{revtex4}

\usepackage{graphicx}% Include figure files
\usepackage{dcolumn}% Align table columns on decimal point
\usepackage{bm}% bold math

\begin{document}

\preprint{APS/es2007apr30-110}

\title{Superconductivity in spinel oxide LiTi$_{2}$O$_{4}$ epitaxial thin films}

\author{Rajesh V. Chopdekar}
\email[Electronic address: ]{rvc2@cornell.edu}
\affiliation{School of Applied and Engineering Physics, Cornell University, Ithaca, NY 14853}
\affiliation{Department of Materials Science and Engineering, University of California, Berkeley, Berkeley, CA 94720}
\author{Elke Arenholz}
\affiliation{Advanced Light Source, Lawrence Berkeley National Laboratory, Berkeley, CA 94720}
\author{Yuri Suzuki}
\affiliation{Department of Materials Science and Engineering, University of California, Berkeley, Berkeley, CA 94720}

\date{\today}

\begin{abstract}
LiTi$_{2}$O$_{4}$ is a unique material in that it is the only known oxide spinel superconductor.  Although bulk studies have demonstrated that superconductivity can be generally described by the Bardeen-Cooper-Schreiffer theory, the microscopic mechanisms of superconductivity are not yet resolved fully.  The sensitivity of the superconducting properties to various defects of the spinel crystal structure provides insight into such mechanisms.  Epitaxial films of LiTi$_{2}$O$_{4}$ on single crystalline substrates of MgAl$_{2}$O$_{4}$, MgO, and SrTiO$_{3}$ provide model systems to systematically explore the effects of lattice strain and microstructural disorder.  Lattice strain that affects bandwidth gives rise to limited variations in the superconducting and normal state properties.  Microstructural disorder such as antiphase boundaries that give rise to Ti network disorder can reduce the critical temperature, but Ti network disorder combined with Mg interdiffusion can affect the superconducting state much more dramatically.  Thickness dependent transport studies indicate a superconductor-insulator transition as a function of film thickness regardless of lattice strain and microstructure.   In addition, surface sensitive X-ray absorption spectroscopy has identified Ti to retain site symmetry and average valence of the bulk material regardless of film thickness.   

\end{abstract}

% PACS, the Physics and Astronomy
                             % Classification Scheme.
 
\maketitle

\section{\label{sec:level1}Introduction}

	Spinel structure oxides offer a wealth of electronic and magnetic ground states across a broad range of temperatures.  Spinel oxides with $3d$ transition metals on the octahedral sites exhibit ferromagnetism, antiferromagnetism, charge ordering, and other types of magnetic and electronic ordering depending on the average valence of the cations.  However, there is only one known oxide spinel superconductor to date, LiTi$_{2}$O$_{4}$ (LTO), with a superconducting phase that persists up to 13 K.  Johnston et al. found superconductivity in LTO as the end member of a solid solution of spinel-structure Li$_{1+x}$Ti$_{2-x}$O$_{4}$ (0$\leq$x$\leq$0.33).\cite{JohnstonMRB, JohnstonJLTP}  In the Li spinels half-integral charge exists on each of the octahedral ions due to the monovalent nature of tetrahedrally coordinated Li.  Unlike the layered structure of superconducting cuprates, LTO has 3D connectivity of edge-sharing TiO$_{6}$ octahedra with average octahedral site valence of $d^{0.5}$ (equal amounts of Ti$^{3+}$ and Ti$^{4+}$).  Oxygen deficient SrTiO$_{3-\delta}$ and LTO both superconduct and both have mixed-valent Ti in octahedral coordination;\cite{DejonghPhysC} however, the Ti$^{3+}$/Ti$^{4+}$ ratio is 1.0 in LTO while there is significantly more Ti$^{4+}$ in SrTiO$_{3-\delta}$.  

	Superconductivity and normal state properties of LTO can be largely explained in terms of a $d$-band superconductor within the Bardeen-Cooper-Schreiffer (BCS) model.\cite{SunPRB, TangPRB}  However, the microscopic mechanisms of superconductivity are still not well understood and open questions still remain as to the role of electron-electron correlations and spin fluctuations in determining some of its properties.  Recent theoretical studies on the composition-driven metal-insulator transition for Li(M,Ti)$_{2}$O$_{4}$ with M=Li and Al \cite{FazilehPRB, FazilehPRL} postulate that electronic correlations are strong in these materials.  Since a quantum site percolation model that reproduces noninteracting behavior does not successfully predict the details of the metal to insulator transition, they conclude that electronic correlations are significant in the LTO parent system as well as Li-excess and Al-doped LTO.
	
	In bulk studies of primarily polycrystalline samples, there have been widely varying normal state and superconducting properties influenced by vacancies,\cite{HarrisonJSSC54_2} Li content, and Ti network disorder.\cite{MoshopoulouJAC,MoshopoulouPhysC}  Studies probing the effects of lattice strain on the superconductivity in bulk LTO have shown the application of hydrostatic pressure to increase the Debye temperature which in turn increases T$_{c}$ as predicted by BCS theory, and such enhancement of T$_{c}$ has been observed in both Li deficient and stoichiometric LTO samples.\cite{LinJLTP, SheltonSSC}  There have been comparatively fewer bulk single crystal studies with systematic characterization of physical properties.  
	
	Epitaxial thin films of LTO are model systems for the systematic study of the role of lattice strain and microstructural disorder on superconducting properties.  Epitaxial lattice strain gives rise to changes in the bandwidth that affect electron-electron correlations in many epitaxial thin film systems.\cite{prellierAPL, AngilellaEPJB, autierPRB}  In addition, systematic variations in epitaxial film microstructural disorder may be obtained through the choice of substrate with lattice parameters not equal to the film lattice parameter due to changes in film growth mode or nucleation of dislocations and other defects.\cite{Henzler96}  Such variation in microstructure can shed light on the various scattering processes that may or may not affect its superconducting properties.  Finally, the choice of substrate orientation may provide insight into either intrinsic or strain-induced anisotropic film properties.\cite{WuPRBanis, HabermeierJoS}   However, to date there has been no report of epitaxial thin film growth, although polycrystalline thin films have been synthesized by Inukai et al.\cite{InukaiTSF94, InukaiTSF128}  
	
	In this paper, we report on the successful synthesis and superconducting and normal state characterization of epitaxial LTO thin films on single crystalline MgAl$_{2}$O$_{4}$, MgO, and SrTiO$_{3}$ substrates.  By comparing the superconducting and normal state transport properties of LTO films on these three different substrates we are able to identify the role of lattice strain, Ti network disorder and stoichiometry on superconductivity.  For example, misregistry and disorder at grain boundaries emerge in films grown on SrTiO$_3$ and MgO due to coalescence of spinel LTO grains that have twice the unit cell of the underlying substrate, and such defects are referred to as antiphase boundaries.  Antiphase boundary disorder has been well-characterized in spinel films grown on MgO substrates via transmission electron microscopy analysis.\cite{HuSuzukiIETM, EerensteinPRB}  These defects disrupt Ti-O-Ti octahedral bond ordering in an analogous manner to the disruption of Fe-O-Fe bond ordering in Fe$_3$O$_4$,\cite{MarguliesPRL, VoogtPRB} and such Ti network disorder would influence carrier transport in LTO films.	On the other hand, Mg interdiffusion combined with antiphase boundary-related Ti disorder has a much larger effect on both normal state and superconducting properties. Thickness dependent transport studies indicate a superconductor-insulator transition as a function of film thickness regardless of substrate.  Surface sensitive X-ray absorption spectroscopy (XAS) has identified Ti to retain site symmetry and average valence of the bulk material.  
	
\section{Experimental Methods}

We have chosen to study epitaxial LTO films on MgAl$_{2}$O$_{4}$ (MAO), MgO, and SrTiO$_{3}$ (STO) substrates.  LTO films on MgO and STO substrates give rise to antiphase boundaries due to the lattice parameter of LTO being twice that of MgO and STO.  These antiphase boundaries provide Ti network disorder in LTO films.  A previous report indicated that diffusion of Mg into LTO could suppress the superconducting phase,\cite{DaltonSynMet} and in this case the source of Mg would be interdiffusion at the film-substrate interface.  Substrates without Mg were used to confirm the trends found in thickness dependence studies on MAO substrates.  LTO films on MAO and STO show minimal interdiffusion at the film-substrate interface while there is significant Mg diffusion on MgO substrates.  Furthermore, each of these substrates placed the film under differing amounts of lattice strain through epitaxy.  

Nominally stoichiometric LTO material prepared by solid-state reaction yields a lattice parameter of 0.8405 nm by powder X-ray analysis.\cite{LambertJSSC75}  Films on MgO (lattice constant a = 0.42112 nm, film-substrate mismatch f = +0.21 $\%$) are under slight tension, while films on MAO (a = 0.80831 nm, f = -3.82 $\%$) are under compressive strain.  Since other spinels have been shown to grow epitaxially on perovskite-structure substrates,\cite{ZhengSci, ChopdekarJEM}  LTO films were also grown on perovskite STO (a = 0.3905 nm, f = -7.07 $\%$).  The perovskite substrate promoted the growth of the spinel superconducting phase in spite of the high compressive strain and anti-phase boundaries due to the unit cell of LTO being twice a perovskite unit cell.  MAO and STO substrates were chosen to perform a more extensive thickness dependence study on both (001) and (110) oriented substrates.

Epitaxial thin films of the normal spinel structure oxide LiTi$_{2}$O$_{4}$ were deposited via pulsed laser deposition on  single crystalline (001) MgO, (001), (111) and (110) MAO, and TiO$_{2}$ - terminated (001) and (110) STO with thickness ranging from 5nm to 1$\mu$m.  In contrast to more equilibrium techniques such as evaporation, pulsed laser deposition can enable growth of phases not stable or difficult to grow in bulk form.  Stability issues in air and the so-called `aging effect' were seen in previous samples of stoichiometric LTO,\cite{InukaiJJAP20_9, XuJLTP} thus the commercial target (Praxair Surface Technologies) was a mixture of the stable phases ramsdellite lithium titanium oxide (Li$_{2}$Ti$_{3}$O$_{7}$)\cite{BuckleyPRB, AbrahamsJSSC} and rutile titanium oxide (TiO$_{2}$) to achieve a Li:Ti ratio of 1:2.  Powder X-ray diffraction confirms the presence of these oxides in the target powder but no spinel-type phase reflections were found.  

Substrate temperatures were held at 450-600 $^{\circ}$C in a vacuum of better than 5x10$^{-6}$ Torr to promote growth of the superconducting phase.  Growth of films at elevated temperatures in 100 mTorr of pure oxygen or at 200 $^{\circ}$C and below in vacuum yielded insulating films with no superconducting transition.  Smooth films with low particulate density (less than 0.05 particles/$\mu$m$^{2}$) were produced with laser fluence of 1-2 J/cm$^{2}$ and a repetition rate of 3 Hz, resulting in a deposition rate of approximately 0.03 nm per pulse.

Rutherford Backscattering Spectrometry (RBS) was used to evaluate both film thickness and composition.  However, quantitative analysis of low-Z ions such as Li and O is difficult, and as such only Ti atom density and uniformity were obtained using this technique. X-ray diffraction both in $\theta$-2$\theta$ and 4-circle mode using Siemens D5000 diffractometers assessed film crystallinity and orientation.  

Soft X-ray absorption spectroscopy (XAS) on the Ti L$_{2,3}$ and O K edges of LTO films was performed at beamlines 4.0.2 \cite{YoungAT} and 6.3.1 \cite{Nachimuthu} of the Lawrence Berkeley National Laboratory Advanced Light Source.  X-ray absorption spectra for LTO films of thickness 15-300 nm were taken at room temperature in both normal and grazing incidences (the sample's surface normal direction was collinear with the incoming photon direction and was tilted 60 degrees from the incoming photon direction, respectively).  Reference spectra were measured on bare substrates as well as powder from the target material.  Spectra were obtained by measuring total electron yield, monitoring the sample drain current as a function of photon energy.  Electron yield detection is surface sensitive with a probing depth of 2-5 nm.  The sample current was normalized to the incoming photon flux as measured using a gold mesh inserted in the beam path.  The degree of linear polarization of the incoming X-ray flux was 99$\pm$1$\%$ for both beamlines.  The lateral dimensions of the X-ray interaction area are much larger than the electron escape depth vertically, thus the measured signal averages over a large number of intragrain volume as well as grain boundaries.  

Normal and superconducting-state magnetic properties were measured in a Quantum Design superconducting quantum interference device (SQUID) magnetometer.  Magnetization measurements were performed with DC field applied both in the plane and out of the plane of the sample. To minimize sample flux trapping when cooling through T$_{c}$, the trapped flux in the SQUID magnetometer's superconducting magnet was reduced to less than 0.28 Oe as calibrated by a 99.9$\%$ pure Dy$_{2}$O$_{3}$ sample.  Transport was measured in a Quantum Design physical property measurement system (PPMS) modified with a Keithley 236 source-measure unit and HP3488A switching matrix.  	Resistivity and Hall measurements were performed both in the normal and superconducting state.  DC and low frequency (f = 13.739 Hz) AC resistivity measurements were carried out in varying fields applied out of the plane of the sample from 2-380 K.  Measurements in both cryostats were taken from room temperature to 2 K in fields of up to 5 T for SQUID magnetometry measurements and up to 7 T for transport measurements.  

\section{Structure}

	Structural analysis indicated that films on all substrates were single-phase and single orientation spinel composition.  X-ray diffraction in $\theta$-2$\theta$ geometry showed spinel phase reflections that were epitaxially matched to the single crystal substrate.  No reflections from constituent phases from the target or polymorphs of TiO$_{2}$ were observed.  Films on MAO and STO under compressive strain exhibited elongation of the out-of-plane lattice parameter.  For example, reciprocal lattice mapping of the film 444 reflection on STO(110) showed almost full relaxation of the film to bulk lattice parameters for film thickness greater than 100 nm, but only partial relaxation below 100 nm.   Careful X-ray diffraction measurements indicated that the lattice parameters of LTO were elongated along the out-of-plane direction with up to 1.7$\%$ elongation from bulk for a 22 nm thick film on (001)MAO.  This result suggests that biaxial compressive strains for ultrathin films are non-volume preserving and therefore may affect the Debye temperature and in turn the T$_{c}$. Films on MgO had reflections which overlay the substrate reflections to within the experimental resolution of the diffractometer, thus the films are under slight tension and show very little contraction of the out-of-plane lattice parameter. 
	
Atomic force microscopy (AFM) indicated the low lattice mismatch between LTO and MgO gave rise to smooth films with an RMS roughness of 0.444 nm (or approximately half of a spinel unit cell) for a 100 nm thick film.  However, rougher film morphology was seen on MAO substrates, with 1.6-2.6 nm RMS roughness for films of similar thickness.  Films on STO had comparable roughness to those on MAO, and film grain size for both substrates at a deposition temperature of 450 $^{\circ}$C or 600 $^{\circ}$C was on average 100 nm. The rougher surface morphology was attributed to the larger epitaxial lattice mismatch and the accompanying full lattice relaxation. Since STO and MgO have approximately half the unit cell size of LTO, we expect antiphase boundaries in the LTO films on STO and MgO but not on MAO. Given the similarities of the STO and MAO samples, these antiphase boundaries appear not to affect the surface morphology. Films on (110) oriented substrates had elongated grains with an aspect ratio of 2:1 favoring the [1$\bar{1}$0] in-plane axis as the fast-growth direction, as well as increased out of plane roughness compared to (001) oriented films.  No measurable anisotropy or roughening compared to (001) films was measured on films grown on (111) oriented MAO, though the average grain size of 30 nm was smaller than grain sizes of (110) films with comparable thickness. 

RBS analysis enabled us to probe the degree of interdiffusion of species at the film/substrate interface. Because of the difficulty in analyzing low Z ions such as Li and O, we focused on analyzing the uniformity of the atomic density of Ti from the LTO as well as the atomic species from the respective substrates. RBS analysis confirmed the interdiffusion of Mg into the LTO films deposited at 600 $^{\circ}$C from MgO substrates with an approximate ratio of Mg to Ti of 0.25:2 assuming a uniform film stoichiometry, but no measurable interdiffusion for films on STO or MAO substrates to within the 5$\%$ accuracy of the measurement.  XAS of Mg K and Al K edges of as-deposited films on MgO (peak to background of 1.25:1 for Mg K edge) as well as films annealed for 10 minutes at 450 $^{\circ}$C in 500 mTorr O$_2$ on MAO (peak to background of 1.07:1 for Mg K edge) confirm interface Mg diffusing from the substrate into the film during growth on MgO when compared to spectra of bare substrates found in the literature.  Similarity of the aforementioned spectra to MgAl$_{2}$O$_{4}$ Mg K edge spectra\cite{BugaevPRB} implies that the interdiffused Mg in LTO/MgO samples substitutes primarily into tetrahedrally coordinated sites.  However, as-deposited films on MAO show little Mg at the surface, with a peak to background of approximately 1.01:1 for the Mg K edge.

\section{Ti L$_{2,3}$ and O K X-ray Absorption Spectroscopy}

\begin{figure}
\includegraphics{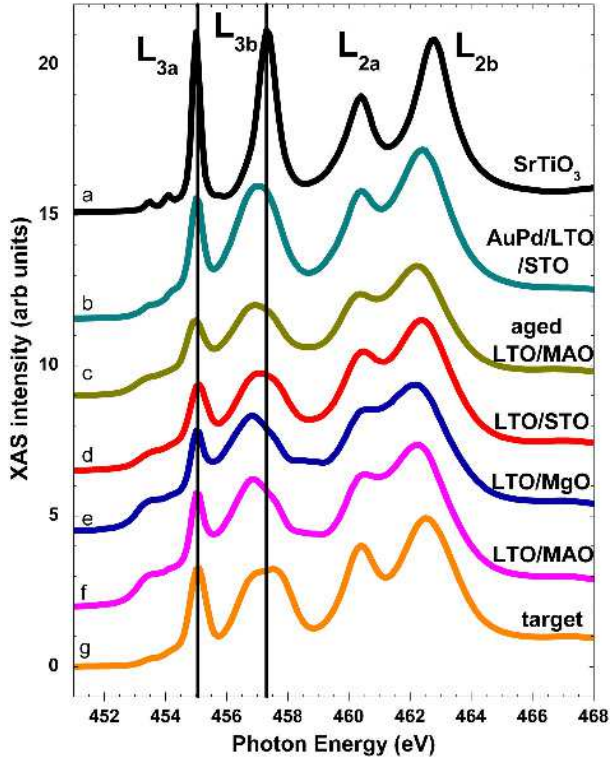}
\caption{\label{TiXASfig}(Color online) Normal-incidence Ti L$_{2,3}$ absorption edge spectra for LTO films on various substrates: (b) 3 nm AuPd / LTO on (001)STO, (c) prolonged air-exposed LTO on (001)MAO, (d) LTO on (001)STO, (e) LTO on (001)MgO, and (f) LTO on (001)MAO, as well as spectra from (a) a bare STO wafer and (g) the pressed powder target mixture as sources of Ti$^{4+}$-only compounds.}
\end{figure}

\begin{figure}
\includegraphics{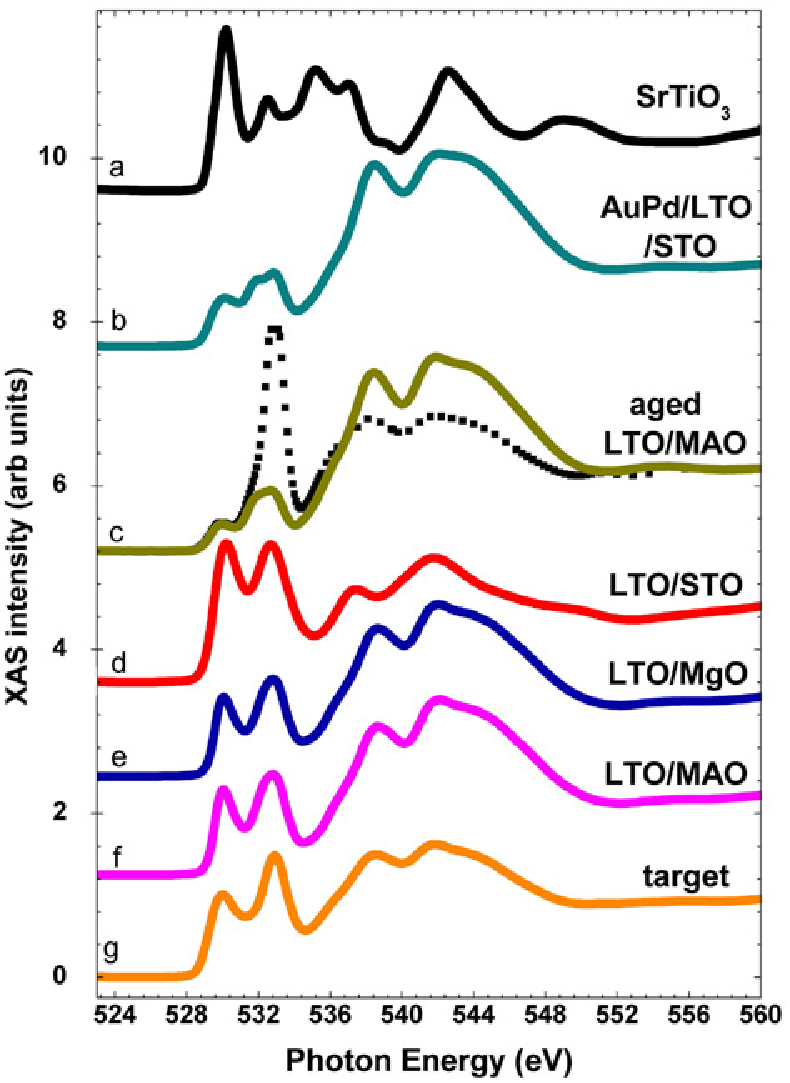}
\caption{\label{OXASfig}(Color online) O K absorption edge spectra for LTO films on various substrates as well as reference spectra as described in Figure \ref{TiXASfig}.  All spectra with solid lines were taken in normal incidence, while the dotted spectra overlaying spectrum (c) was taken in grazing incidence.}
\end{figure}

In order to probe the cation environment and its effect on the observation of superconductivity, surface sensitive soft X-ray absorption spectroscopy was performed at the Ti L$_{2,3}$ and O K absorption edges on films of varying thickness below 500 nm to determine the Ti ion environment.  Spectra have been aligned to the first sharp peak at each absorpton edge: 455 eV for the Ti L$_{3a}$ peak (Figure \ref{TiXASfig}) and 530 eV for the first O K edge peak (Figure \ref{OXASfig}).  Unlike heavier $3d$ transition metals like Fe with two dominant features at the L$_{2,3}$ edge, each of the Ti L$_{3}$ (454-458 eV) and L$_{2}$ (460-465 eV) absorption features are split into qualitatively t$_{2g}$ and e$_{g}$-like sub-peaks.  Crystal field effects have a large effect on the relative intensity of each of these peaks for the case of Ti$^{4+}$ in different environments.\cite{deGrootPRB, deGrootPCM}  For the case of Ti$^{3+}$ in compounds such as LaTiO$_{3}$, the lower energy t$_{2g}$-like peaks have low intensity compared to the e$_{g}$-like peaks.  Since bulk LTO has an equal number of Ti$^{3+}$ and Ti$^{4+}$ in octahedral environments, one would expect a spectrum similar to mixed-valence octahedral Ti such as in La$_{1-y}$Sr$_{y}$TiO$_{3}$.\cite{AbbatePRB}  Comparison of the STO Ti$^{4+}$-only spectrum in Figure \ref{TiXASfig} (a) to spectra (d)-(f) shows that LTO spectra have an increase in spectral weight at a 454 eV pre-peak feature at the expense of L$_{3a}$  intensity, as well as a merging of the L$_{2a}$ and L$_{2b}$ peaks.  This trend is also seen in the y=0.4 and y=0.6 spectra from Abbate et al's study on La$_{1-y}$Sr$_{y}$TiO$_{3}$.\cite{AbbatePRB}  

Ra et al.\cite{RaAPL04} examined powders of stoichiometric and Li-excess Li$_{1+x}$Ti$_{2-x}$O$_{4}$ and found qualitatively similar Ti L$_{2,3}$ lineshapes for 0$\leq$x$\leq$0.33.  Following their analysis, Lorentzian fits were made simultaneously to the pre-edge feature at 454 eV as well as each of the L$_{3a}$, L$_{3b}$, L$_{2a}$, and L$_{2b}$ peaks.  The ratio of L$_{3a}$ to L$_{3b}$ peak areas was approximately 0.14 for LTO on MgO and MAO, and showed a trend of increasing L$_{3a}$ contribution for films on STO and aged samples.  This suggests that the surface of aged samples as well as the those of STO have slightly more Ti$^{4+}$ character than similar films on MgO or MAO.  Samples on STO capped with 3 nm AuPd or SrRuO$_{3}$ deposited in-situ at 400 $^\circ$C showed similar spectra to uncapped samples, suggesting that the more Ti$^{4+}$-like spectrum is intrinsic to the surface and interface of as-deposited LTO films on STO.  

In contrast, the O K edge features may be divided into two regions: two low-energy peaks at 530 eV and 532.5 eV that show strong hybridization between O $2p$ and Ti $3d$ states, and a broader region between 536-548 eV exhibiting mixing of Ti $4sp$ and O $2p$ states.\cite{deGrootPRB}  The lower-energy peaks for the various samples resemble the spectrum from the target material due to similar Ti-O hybridization with the intensity for the two peaks for the fresh samples equal in magnitude and insensitive to film thickness from 15 nm to 87 nm on STO.  Since these lower-energy peaks are sensitive to neighboring Ti ions, one would expect similar spectra between the edge-sharing TiO$_{6}$ octahedra in LTO and the distorted edge-sharing octahedra in ramsdellite Li$_{2}$Ti$_{3}$O$_{7}$ and rutile TiO$_{2}$ when compared to the corner-sharing octahedra in STO.  One would also expect dramatic changes between fresh and aged sample spectra, but the aged LTO/MAO sample (Figure \ref{OXASfig} (c)) showed large changes only when measured in grazing incidence.  Grazing incidence measurements reduce the effective X-ray penetration depth by cos($\theta$), thus a very shallow volume is probed in grazing incidence.  A similar but weaker trend of suppression of the 530 eV peak when compared to the 532.5 eV peak can be seen in the AuPd capped LTO/STO sample (Figure \ref{OXASfig} (b)) as the AuPd cap also serves to reduce the effective probing depth of the LTO underlayer, but also protects the LTO material from reaction with air.  Together these imply that while aging can affect the surface of LTO samples dramatically, the depth of aged material is on the order of the photoelectron escape depth (of order 5nm\cite{FrazerSS03} but is material-dependent).

Again, comparison to the spectra from Ra et al. confirm that the surface material is close to target stoichiometry as-deposited and such surface material decreases in Li content as the sample ages.  Aging for 48 hours in air affects the surface-sensitive XAS lineshape, but only with aging for longer timescales do significant changes in normal and superconducting state transport properties occur.  Therefore the variations in transport properties for different samples described below cannot be attributed to variations in Ti average valence, Ti site symmetry or Li deficiency and hence Ti network disorder.

Although XAS spectra can show strong differences based on site symmetry for isovalent compounds such as rutile and anatase TiO$_{2}$,\cite{deGrootPCM} there is no clear distinction as a function of thickness for the LTO films shown in Figures \ref{TiXASfig} and \ref{OXASfig}.  Thus the average valence and site occupation for the surface monolayers of each film should be identical, in spite of the presence of measurable superconductivity in thicker films and no resistive superconducting transition in a 15 nm film on STO.  Thick and relaxed films on MAO, MgO and STO substrates also have similar features to the thin films on STO, indicating that the surface monolayers do not change substantially post-deposition as a function of substrate, and the XAS surface measurement is insensitive to the presence of anti-phase boundary disorder. 

\section{Magnetism}

\begin{figure}
\includegraphics{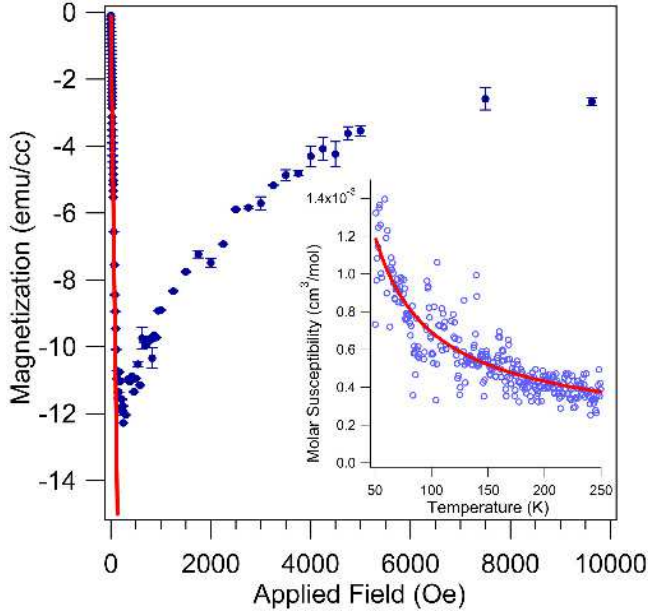}
\caption{\label{fig1}(Color online) Main: Low-field diamagnetic response at 1.8 K of a zero-field cooled 350 nm LTO film on (001)MAO with magnetic field in the plane of the sample.  Deviation from linearity takes place at 46$\pm$3 Oe.  Inset: molar susceptibility of a 400 nm LiTi$_{2}$O$_{4}$ film on (001)STO measured with an in-plane field of 50,000 Oe.  A fit to the data is indicated as a solid line (see text).}
\end{figure}

The magnetic response of LTO films was measured on all substrates and found to be comparable to LTO bulk single crystals, polycrystalline pellets, and powder samples.  Zero-field cooled samples at 1.8 K show diamagnetic shielding with low applied fields.  A linear extrapolation at low applied fields for the data presented in Figure \ref{fig1} yielded a typical lower critical field H$_{c1}$ of 46$\pm$3 Oe at which point the diamagnetic response deviated from linearity by 1 $\%$.  

In order to understand the factors governing the normal state response of our films, we also measured the normal state susceptibility and fit it to a modified Curie-Weiss law.\cite{JohnstonJLTP}  The Curie-Weiss behavior can result from trapped electrons on defect sites as well as impurities, whereas the weakly temperature dependent residual susceptibility $A_{m} + B_{m}T$ in the model below is due to Pauli paramagnetism from conduction electrons as well as response from Ti$^{3+}$-Ti$^{3+}$ dimers.  
\begin{equation}
\chi(T) = \frac{C_{m}}{T-\theta} +  A_{m} + B_{m}T
\end{equation}
Figure \ref{fig1} inset shows the normal state susceptibility of a 400 nm LTO film on an STO substrate with the substrate background signal subtracted.  A least-squares fit to the measured film susceptibility produced Curie constant C$_{m}$ = 0.0398$\pm$0.001 cm$^{3}\cdot$K/mol and temperature $\theta$ = 13.6$\pm$2.1 K if the fit is constrained to match bulk values of A$_{m}$ = 0.00225 cm$^{3}$/mol, and B$_{m}$ = -9x10$^{-8}$ cm$^{3}$/mol$\cdot$K for stoichiometric LTO.\cite{HarrisonPhilMag85, HarrisonJSSC54_2}  A large uncertainty in the fit parameters for the temperature independent coefficient A$_{m}$ and linear temperature-dependent coefficient B$_{m}$ are present if all parameters are left floating due to the film signal approaching the DC SQUID magnetometry noise floor of order 1x10$^{-7}$ emu at 50,000 Oe.   Localized electrons at oxygen defects have been shown to increase the paramagnetic response of LTO, and an upper limit of 10 $\%$ of such impurities are present in the film assuming only trapped electrons of spin 1/2 are responsible for the entirety of the Curie-Weiss behavior.\cite{HarrisonJSSC54_2}  For our LTO films which were grown in a reducing environment, the expected amount of oxygen defects per sample volume was larger than had been seen in previous bulk studies. In fact, the much larger surface to volume ratio and small grain size pointed to non-stoichiometry at the film surface and at grain boundaries.  Accordingly, the Curie constant extracted from our DC susceptibility measurements showed an increase over that of previously measured bulk samples but were similar for our films on different substrates. 

\section{Transport}\label{trans}

In order to probe the effects of lattice strain and microstructural disorder on the superconducting transition, we performed resistivity and Hall effect measurements on our LTO films.  In particular, we discuss the normal state resistivity values, resistive transition temperatures and widths, the upper critical field, the Ginzburg-Landau coherence length-mean free path product and Hall mobility as a function of substrate and film thickness.

\begin{figure}
\includegraphics{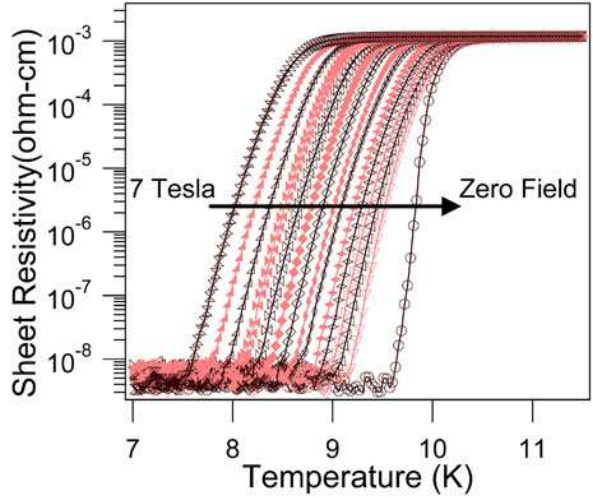}
\caption{\label{MAOResVsHT}(Color online) Zero-field cooled field-dependent resistivity taken on warming as a function of temperature of a 60 nm film on (001)MAO in the temperature regime around its superconducting transition.  Measurements were taken with field applied out of plane at every 0.5 T from 7 T to zero field, with bold lines indicating data at 1 T intervals.}
\end{figure}

  The normal-state resistivities versus temperature for nominally stoichiometric films on MAO (e.g. Figure \ref{MAOResVsHT}) and STO deposited at 600 $^{\circ}$C are comparable to that of polycrystalline thin film samples\cite{InukaiTSF94} as well as bulk polycrystalline samples\cite{UedaJSSC} despite the presence of large compressive epitaxial strains.  The film on (001) MAO described in Figure \ref{MAOResVsHT} as well as a film of comparable thickness on (001) STO both have a resistivity of 1.2x10$^{-3} \Omega \cdot$cm at 12K.  The similarity in magnitude of normal-state resistivity suggests that the presence of anti-phase boundaries, and hence Ti network disorder, in films on STO does not have a significant effect on the normal state transport in such LTO films.  In contrast, films on MgO were found to have an order of magnitude greater normal-state resistivity despite having minimal epitaxial strain and much smoother film morphology compared to films on STO or MAO. The significantly larger normal state resistivity values suggest that partial Mg$^{2+}$ interdiffusion into octahedral sublattice sites\cite{LambertJSSC75, DaltonSynMet, SteinbruckJMSL} at the 600 $^{\circ}$C deposition temperature coupled with antiphase boundary-related Ti network disorder\cite{MoshopoulouPhysC} give rise to greater scattering in the normal state.  A lower deposition temperature of 450 $^{\circ}$C yielded films with higher residual resistivity, though superconducting transitions for films on MAO and STO remained at approximately 10 K.  Interdiffusion does not play as large a role at a 450 $^{\circ}$C deposition temperature as confirmed by an increase in the Mg K XAS edge intensity in films grown at 600 $^{\circ}$C.  Thus, the higher residual resistivity implies that scattering from defects such as anti-phase boundaries is increased for a given film thickness.  
     	  
\subsection{Superconducting-State Transport} \label{sst}

\begin{figure}
\includegraphics{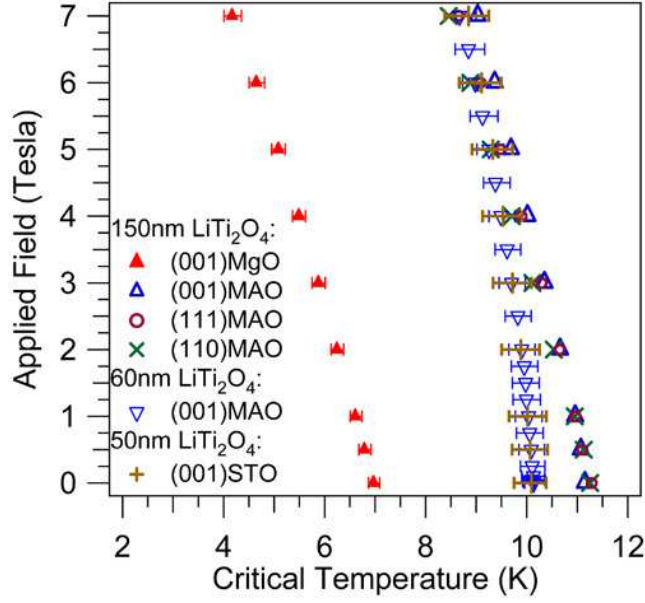}
\caption{\label{CtvsHc2}(Color online) Critical temperature versus applied field for 50 nm, 60 nm and 150 nm LTO films on various substrates with magnetic field applied out of the plane of the sample.  The horizontal error bars indicate the superconducting transition temperature width (10$\%$-90$\%$ of the resistivity at 15 K).}
\end{figure}

	Figure \ref{CtvsHc2} plots a summary of the resistive transition temperatures and widths measured for LTO films of varying thicknesses grown on different substrates and substrate orientations. Mirroring the trend observed with the normal state resistivity values, the films grown on (001) MAO and STO display similar behavior with a sharp zero field transition width (for 10$\%$ to 90$\%$ of normal state resistivity at 15 K) and transition temperature near 10.8 K.  Nominally unstrained films on (001) MgO show a sharper zero field transition width of 0.3 K but with transition temperatures depressed to 6.9-9 K.  The lower transition temperatures on MgO substrates, in contrast to those on MAO and STO substrates, can result from more significant Mg interdiffusion into the LTO film.\cite{LambertJSSC75}  Films on (110) or (111) oriented MAO substrates show broader transition widths of up to 1.5 K which may be linked to smaller grain size in AFM scans.  The increased number of non-superconducting grain boundaries in such samples may dominate the resistivity measurement, or the intragrain volumes themselves may be off-stoichiometry.  Similar broadening of transition widths without significant degradation of the transition temperature were seen in powder pellets of LTO\cite{HarrisonPhilMag85} as well as artificial YBa$_{2}$Cu$_{3}$O$_{7}$ superlattices.\cite{NortonPRL}  The broadening in the aforementioned studies was not due to intrinsic properties of the superconducting regions but instead the boundary regions with the non-superconducting material.  Since the LTO film transition temperature is not suppressed as a function of orientation for the 150 nm thick LTO films in Figure \ref{CtvsHc2}, this suggests that grain boundaries rather than whole-film non-stoichiometry is the source of the transition broadening.  If the grain boundary resistance in LTO films can be a significant fraction of the measured resistance, careful analysis must be performed to measure intrinsic LTO properties from boundary effects as we will discuss in Section  \ref{nstthick}.  Finally, films on STO have comparable transition temperatures to films on MAO, implying that antiphase boundaries have little effect on the magnitude of the critical temperature.

A closer look at the upper critical field and Ginzburg-Landau coherence length-mean free path product of LTO thin films on different substrates indicates the presence of a disordered phase either at the surface or film-substrate interface in thinner LTO samples.  Using the Werthamer, Helfand and Hohemberg \cite{WHH_PR} (WHH) model for a type II superconductor in the dirty limit, we can estimate the upper critical field ($H_{c2}$) from the low-field slope of the critical temperature as a function of applied magnetic field.  The extracted  $H_{c2}$(T=0 K) values ranged from 12-24T for twenty-five samples depending on substrate choice and orientation, and the coherence length $\xi(T=0 K)$ may be estimated from the Ginzburg-Landau formula $H_{c2}=\Phi_{0}/2\pi\xi^{2}$ as 3.5-5.5 nm which is consistent with bulk values.  Foner and McNiff\cite{FonerSSC} found that in spite of different starting compositions of Li$_{2.6}$Ti$_{2-z}$O$_{4}$, T$_{c}$ values were uniformly 12 K for stoichiometry deviations in the range of $-0.7<z<0.5$.  However, there was strong variation in the quantities (d$H_{c2}$/dT)$_{T=T_{c}}$ and $\lambda_{so}$, the zero-field slope of the $H_{c2}$ vs T$_{c}$ curve and the spin-orbit scattering parameter, respectively.  In analogy, the T$_{c}$ of films on MAO and STO for a given thickness is suppressed in thinner films to a value of on average 10.8$\pm$0.5 K compared to 900 nm thick films with a critical temperature of 11.3 K.  Thinner films such as the 60 nm film on (001)MAO or 50 nm film on (001)STO in Figure \ref{CtvsHc2} have suppressed critical temperatures in zero field, though in high field their behavior approaches those of thicker films. 
 On the other hand, the films on MgO had a suppressed critical temperature but fit very well to the universal WHH curve when the reduced field $h_{c2}$ = $H_{c2}$(T)/[T$_{c}$(H=0)$\cdot$(d$H_{c2}$/dT)$_{T=T_{c}}$ ] is plotted as a function of reduced critical temperature t=T$_{c}$(H)/T$_{c}$(H=0).  The increase in the reduced $h_{c2}$ for low applied fields over that estimated from the WHH model can result from increased localization due to a disordered or inhomogeneous phase \cite{CoffeyPRL, TenhoverSSC} in thinner LTO samples, whereas the significant interdiffusion of Mg in films on MgO yields a more uniform Li$_{u}$Mg$_{v}$Ti$_{2}$O$_{4}$ phase.  

The Ginzburg-Landau coherence length-mean free path product $\xi_{0}l$ may also be extracted from the transport data using the dirty-limit formula
\begin{equation}
\xi(T)=0.855 \Bigl(\frac{ \xi_{0}l}{1-T/T_{c}}\Bigr)^\frac{1}{2} 
\end{equation} 
For the resistive transitions plotted in Figure \ref{CtvsHc2}, $\xi_{0}l$ for the 150 nm thick films is close to 1400 \AA$^{2}$ and is consistent with crystals at 5$\%$ or closer in composition to the stochiometric LiTi$_{2}$O$_{4}$ phase.\cite{UedaJSSC, SunPRB}  However, the 50 nm and 60 nm films on both MAO and STO have $\xi_{0}l$ values close to 800 \AA$^{2}$, suggesting that deviations from bulk-like behavior at either the surface or the film-substrate interface dominate at these thicknesses.

\subsection{Normal-State Transport, $t \geq 100nm$}\label{nstthick}

The Hall mobility of films on MAO and STO substrates was measured by computing the ratio between the Hall resistance and the sheet resistance in the van der Pauw configuration from the critical temperature up to room temperature.  Assuming a simple one-carrier type model, films are n-type with carrier concentration of 1.3x10$^{22}$ cm$^{-3}$ for most films; this carrier concentration is equivalent to a single carrier per LTO formula unit as one would expect from the mixed-valent Ti ions.  Carrier concentration and mobility were extracted by simultaneous van der Pauw and Hall measurements as a function of applied field out of the plane of the sample up to 7T between 11 K and 305 K.\cite{vanderpauw58L}  

Films deposited on STO and MAO at higher temperatures (600 $^\circ$C) showed better crystallinity as confirmed by XRD rocking curves as well as mobility as high as 0.5 cm$^{2}$/V$\cdot$s at room temperature as measured by the Hall effect.  Similar values for LTO mobility were found for polycrystalline sputtered LTO films annealed at 850 $^\circ$C.\cite{InukaiTSF128}  A suppression of mobility by a factor of 2 was seen in films deposited on STO at 450 $^\circ$C, but the carrier concentration remained at 1 electron per LTO formula unit for films on STO at both 450 $^\circ$C and 600 $^\circ$C. 

\begin{figure}
\includegraphics{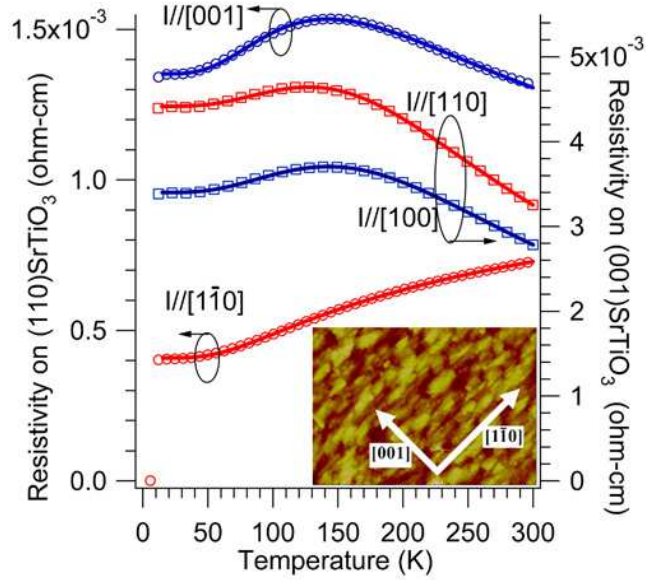}
\caption{\label{fig4}(Color online) Resistivity of thin rectangular LTO film sections along different crystallographic directions with every 15th point plotted for (001)STO (squares) and (110)STO (circles).  Each data set indicates along which crystallographic direction current was directed during measurement.  Solid lines indicate fits to equation \ref{eq_twocomprho} as detailed in the text.  Inset: AFM scan of 1.75 x 2.5 micron area on an LTO film on (110)STO showing elongation of grain structure along the [1$\bar{1}$0] substrate direction.}
\end{figure}

A detailed look at the normal state resistivity of LTO shows non-monotonic behavior with both positive and negative slope.  Such features have been attributed to non-stoichiometry or substitution in LTO bulk samples.\cite{SteinbruckJMSL, TangPRB}  As discussed in Section \ref{sst}, grain boundary resistance may play a role in the measured sample resistance and thus must be taken into consideration when performing quantitative analysis on transport data.  Thus, a two component model of resistivity consisting of metallic and semiconducting components was used to account for disorder and inhomogenieties in our thinner superconducting LTO films as well as films deposited at 450 $^{\circ}$C.  Similar modeling was successfully applied to (Ba,K)BiO$_{3}$ films.\cite{HellmanPRB} 
\begin{equation}
\label{eq_twocomprho}
\rho(T)=
\Biggl\{
(1-A)\biggl[
\frac{1}{\rho_{s}e^{E_{s}/k_{B}T}}
\biggr]+
\frac{A}{\rho_{m}(T)}
\Biggr\}^{-1} 
\end{equation} 

The first bracketed term models the semiconducting component of the resistivity which freezes out at low temperatures, and is parametrized by the semiconducting channel resistivity asymptote $\rho_{s}$ and hopping activation energy $E_s$.  The second term models a metallic resistivity, and the weighting parameter $A$ details the relative contributions between the metallic and semiconducting channels.  The normal-state resistivity has been fit in vanadium-silicon alloy films\cite{NavaPRB86} using the electron-phonon scattering form of the Bloch-Gruneisen equation with a series residual resistivity $\rho_{0}$ as well as a phenomenological parallel saturation resistivity $\rho_{sat}$ (equation \ref{eq_rhom}).  The Bloch-Gruneisen equation (equation \ref{eq_bg}) parameters include the high temperature resistivity coefficient $\alpha_{BG}$ and the Debye temperature $\theta_{D}$.  We may use this model for the metallic resistivity term in equation \ref{eq_twocomprho} for samples with non-monotonic resistivity, as well as only using equation \ref{eq_rhom} for samples with monotonic metallic resistivity such as 60 nm and thicker films on both MAO and STO substrates deposited at 600 $^{\circ}$C.

\begin{equation}
\label{eq_rhom}
\frac{1}{\rho_{m}(T)} =  \biggl[\frac{1}{\rho_{sat}} + \left\{ \rho_{0}
+ \rho_{BG}
\right\}^{-1}\biggr]
\end{equation}
\begin{equation}
\label{eq_bg}
\rho_{BG} = \frac{4\alpha_{BG}T^5}{\theta_{D}^4}\int^{\theta_{D}/T}_{0}{dz\frac{z^5}{(e^{z} - 1)(1 - e^{-z})}}
\end{equation}

For samples that have monotonic resistivity with a positive temperature coefficient of resisitivty, such as the LTO/(110)STO I//[1$\bar{1}$0] data in Figure \ref{fig4}, equation \ref{eq_rhom} may be used to obtain $\alpha_{BG}$ and $\theta_{D}$.  For a range of different sample thicknesses with non-monotonic resistivity, equation \ref{eq_twocomprho} is better suited to fit the data by leaving $A$, $\rho_{s}$, $E_s$, $\rho_{sat}$, and $\rho_{0}$ floating during data fitting while constraining $\alpha_{BG}$ and $\theta_{D}$ to match values obtained from samples with monotonic resistivity.

Values for the LTO/(110)STO I//[1$\bar{1}$0] data in Figure \ref{fig4} when fit to equation \ref{eq_rhom} are $\alpha_{BG}$ = 4.6x10$^{-6}  \Omega \cdot$cm/K and $\theta_{D}$ = 420 K.  The resulting Debye temperature of 420 K is suppressed compared to measured Debye temperature of 537-700 K via heat capacity experiments.\cite{SunPRB}  Tang et al.\cite{TangPRB} assumed a closed-form solution of the Bloch-Gruneisen equation by assuming an Einstein phonon distribution, and their best fit Einstein temperature of 367 K is consistent with our best fit temperature of approximately 280 K with the same closed-form equation applied to the data in Figure \ref{fig4}.

If one compares the films on STO from Figure \ref{fig4} and their twins on (001) and (110)MAO using the dimensionless weighting parameter $A$, one may evaluate the relative contribution of the metallic and semiconducting components in the fit.  Elongation of the grain structure for films on (110)STO (e.g. Figure \ref{fig4}) and MAO allows for the measurement of differing ratios of intragrain and intergrain volumes along different crystallographic directions in the same sample.  Thus by using equation \ref{eq_twocomprho} one may separate the intrinsic LTO resistivity from contributions to the resistivity by non-intrinsic sources such as grain boundaries and other microstructural disorder.  As the grains on (110)STO are elongated along the [1$\bar{1}$0] direction, current confined along the [1$\bar{1}$0] direction probes a large fraction of intragrain volumes with fewer intergrain areas contributing to the resistivity.  On the other hand, when current is confined along the [001] in-plane direction for the same sample, the relative amount of intergrain resistivity contributions are increased, and accordingly the measured sheet resistivity increases as well as a decrease in resistivity with increasing temperature above 150K.  This suggests that while intragrain volumes are well-connected along the [1$\bar{1}$0] direction, grain boundary resistance can have substantial influence along the [001] crystallographic direction.  The similarity in shape for the (001)STO resistivity data along both the [100] and [110] in-plane directions point to a more isotropic contribution of grain boundary resistance on the measured sheet resistivity.

For the above samples, $A$ falls between 1.0 (LTO on (110)STO with current along [1$\bar{1}$0]) and 0.16 (LTO on (001)STO with current along [110]) suggesting that even in thick films the semiconducting channel is not negligible.  Additionally, $A_{110} \approx A_{100}$ for films on STO and MAO confirms that there is little anisotropy in the ratio of the two resistivity channels along different crystallographic directions for (001) oriented films.  Finally, $A_{MAO} > A_{STO}$ indicates that antiphase boundaries in films on STO can increase the apparent contribution from a high resistivity channel in the temperature region of 20-100K for thick LTO films without substantially affecting the T$_c$.

\subsection{Normal-State Transport, $t \leq 100nm$}

\begin{figure}
\includegraphics{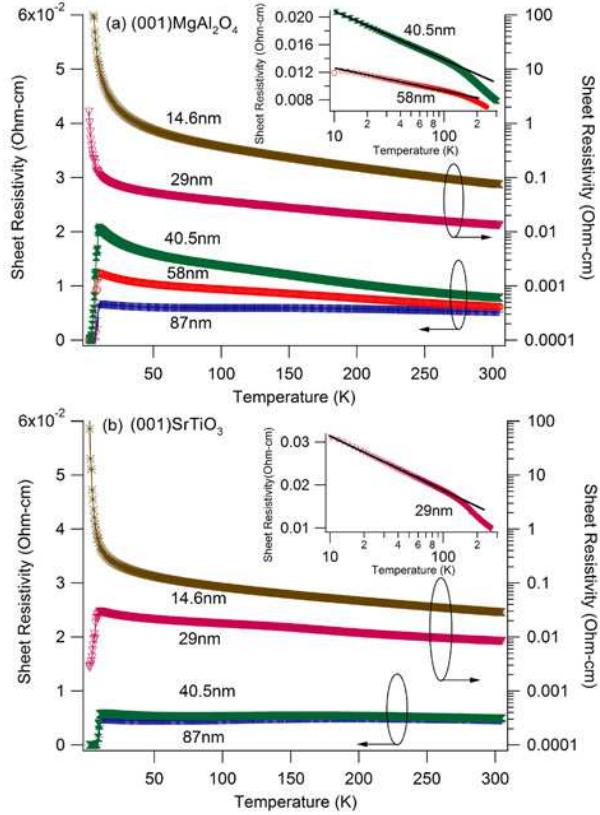}
\caption{\label{ResThicknesTrend}(Color online) Main: Sheet resistivity of thin LTO films deposited on (a) (001)MAO and (b) (001)STO substrates at a growth temperature of 450 $^{\circ}$C.  Below 15 nm, films on both substrates exhibit suppression of the superconducting state as well as an increase in resistivity above 15K over that of thicker films.  Insets: Sheet resistivity as a function of temperature on a logarithmic scale for selected samples with lines as guides to the eye.}
\end{figure}

	While films thicker than 100 nm on MAO and STO showed similar transport and magnetization characteristics, there was a distinct thickness dependence on the transport for films thinner than 100 nm.  Figure \ref{ResThicknesTrend} shows the normal state resistivities for LTO films on MAO and STO deposited at 450 $^{\circ}$C with thickness ranging from 14.6 - 87 nm.  The resistive superconducting transition in LTO films disappears for 29 nm thick LTO films on MAO and 14.6 nm thick films on STO. 
	
	The disappearance altogether of a superconducting transition in very thin LTO films is a signature of a superconductor-insulator transition as a function of film thickness.  This transition has been seen in other systems and originated from Coulomb interactions or impurity scattering that localizes states.\cite{BergmannPRB, JaegerPRB86}  The onset of insulating behavior occurs for thicker films in the case of LTO/MAO samples compared to those of LTO/STO, thus suggesting that the impurity scattering from interfacial defects such as misfit dislocations\cite{MatthewsMD} due to epitaxial strain between the film and the substrate in partially strained LTO/MAO are more effective in the suppression of superconductivity than misfit dislocations and antiphase boundaries in relaxed LTO/STO samples.   However, with the exception of the thinnest films, the similarity in field dependent critical temperatures  (Figure \ref{CtvsHc2}) indicates that such defects do not, for the most part, impact the superconducting properties.

A two component resistivity model may also be applied to LTO films thinner than 100 nm with high residual resistivity.  However, in lieu of the Bloch-Gruneisen form of the metallic resistivity component, a logarithmic resistivity is better suited to the temperature regime between 15 K and 100 K (Figure \ref{ResThicknesTrend}, insets).  Such a logarithmic dependence on temperature signifies either weak localization of carriers or defect and impurity-induced electron scattering as seen in granular or amorphous metal films.\cite{DeutscherPRL, BergmannPRB}  A strong magnetoresistance is found for systems where weak localization is dominant,\cite{BergmannPRB} but a weak magnetoresistance at 5 T of less than 0.1 \% points to a defect-induced scattering mechanism such as antiphase boundary scattering as the source of logarithmic temperature dependence on resistance.  As this logarithmic behavior is seen clearly in films on STO deposited at 450 $^{\circ}$C ($e.g.$ Figure \ref{ResThicknesTrend}(b) $t=$29 nm) but a Bloch-Gruneisen type fit is more suitable for films of the same thickness on STO deposited at 600 $^{\circ}$C, the higher deposition temperature seems to reduces antiphase boundary defect density such that this scattering channel is minimized.  Previous studies on films with antiphase boundaries have shown that high-temperature annealing results in grain growth and a reduction in antiphase boundary density.\cite{VenzkeJMR}  A further crossover as a function of thickness from a logarithmic dependence to exponential temperature dependence of sheet resistance and stronger localization occurs for the thinnest films ($e.g.$ Figure \ref{ResThicknesTrend} (a) or (b) $t = $14.6 nm), and this normal-state resistance crossover is accompanied by a subsequent complete suppression of the superconducting transition.  Such a crossover is seen in superconducting metal films\cite{DeutscherPRL, JaegerPRB89} at a normal-state sheet resistance around the value of the pair breaking resistance $R_{sq}=\frac{h}{4e^2}=6.4k\Omega$.  This crossover occurs for LTO film sheet resistance values in the range of $3-8k\Omega$ as determined by a linear extrapolation of the suppression of T$_c$ as a function of sheet resistance\cite{GraybealPRB84} of films both on MAO and STO.  
 
\section{Conclusion}

Epitaxial films of the spinel superconductor LiTi$_{2}$O$_{4}$ were grown on a variety of substrates to explore the effects of strain and microstructure on measured normal state and superconducting properties.  Initial transport and magnetic measurements were consistent with bulk samples with an average critical temperature of 10.8 K and sharp transition width for films deposited on (001)MgAl$_{2}$O$_{4}$ and SrTiO$_{3}$. The surface properties of freshly-deposited films on a variety of substrates were identical as measured by soft X-ray absorption spectroscopy on the Ti L$_{2,3}$ edges, though substantial aging of the LTO samples can show a spectroscopic signature in the O K edge spectrum low energy 530 eV and 532.5 eV peaks.  Detailed transport evaluation indicated that measured properties of the films were composed of contributions from metallic intragrain regions weighted with semiconducting grain boundary regions.  Anti-phase boundaries have a strong anomalous effect on the magnetization of spinels oxides such as Fe$_{3}$O$_{4}$, but little if any contribution of this type of defect alone was seen when comparing the normal and superconducting transport and magnetic properties of  films thicker than 100 nm deposited on perovskite SrTiO$_{3}$ and isostructural spinel MgAl$_{2}$O$_{4}$-type substrates.  By contrast, the interface Mg interdiffusion combined with antiphase boundaries in LTO films on MgO substrates gives rise to suppressed critical temperature accompanied by higher than bulk normal-state resistivity values.  Comparison of the normal and superconducting state properties of LTO films on all three substrates indicate that lattice strain and hence bandwidth changes do not have a significant effect on the superconducting or normal state properties. Together these results indicate the robustness of the superconducting state of LiTi$_{2}$O$_{4}$ to lattice strain and microstructural disorder.

\begin{acknowledgments}
The authors thank Prof. A. Stacy (UC Berkeley, Chemistry) for the use of her $\theta$-2$\theta$ diffractometer and K. M. Yu (Lawrence Berkeley Lab, Materials Science Division) for taking RBS spectra.  Additional thanks to J.S. Bettinger and B. B. Nelson-Cheeseman for assistance in collecting XAS spectra.  Thanks to Y. Matsushita, I.R. Fisher and M.R. Beasley (Stanford Univ, Applied Physics) for fruitful discussion, and Y. Takamura (University of California, Davis, Chemical Engineering and Materials Science) for insightful comments.  This research was supported by the Office of Naval Research (N00014-97-1-0564) managed by Dr. Colin E.C. Wood.  The Advanced Light Source is supported by the Director, Office of Science, Office of Basic Energy Sciences, of the U.S. Department of Energy under Contract No. DE-AC02-05CH11231.
\end{acknowledgments}

\end{document}